\documentclass[amsmath,amssymb,aps,prb,twocolumn,floatfix,10pt,superscriptaddress]{revtex4-2}

\usepackage{mathtools}
\usepackage[caption = false]{subfig}
\usepackage{graphicx}
\usepackage{standalone}
\usepackage{dcolumn}
\usepackage{bm}
\usepackage{xcolor}
\usepackage{physics}
\usepackage{enumitem}
\usepackage{calligra}
\usepackage[colorlinks = true,linkcolor = red,citecolor = magenta]{hyperref}
\usepackage[sort&compress]{natbib}
\usepackage{orcidlink}
\usepackage[normalem]{ulem}
\usepackage{pdfpages}

\makeatletter
\AtBeginDocument{\let\LS@rot\@undefined}
\makeatother

\DeclareGraphicsExtensions{.pdf,.eps,.png,.jpg,.mps}

\newcommand{\ctiqsadd}{Center for Trapped Ions Quantum Science, Institute for Basic Science, Daejeon 34126, Republic of Korea}
\newcommand{\pcsadd}{Center for Theoretical Physics of Complex Systems, Institute for Basic Science (IBS), Daejeon, Korea, 34126}
\newcommand{\ustadd}{Basic Science Program, Korea University of Science and Technology (UST), Daejeon 34113, Republic of Korea}

\newcommand{\sect}[1]{\emph{#1}\,---\,}

\makeatletter
\renewcommand*{\fnum@figure}{{\normalfont\bfseries \figurename~\thefigure}}
\renewcommand*{\@caption@fignum@sep}{\textbf{:}}
\makeatother

\newcommand{\mc}{\mathcal}

\DeclareMathOperator*{\argmin}{argmin}

\newcommand{\appropto}{\mathrel{\vcenter{
  \offinterlineskip\halign{\hfil$##$\cr
    \propto\cr\noalign{\kern2pt}\sim\cr\noalign{\kern-2pt}}}}}

\newcommand{\dgap}{\ensuremath{\Delta_\mathrm{gap}}}
\newcommand{\heff}{\ensuremath{\mathcal{H}_\mathrm{eff}}}
\newcommand{\neff}{\ensuremath{\mathcal{N}_\mathrm{eff}}}
\newcommand{\leff}{\ensuremath{\mathcal{L}_\mathrm{eff}}}

\newlength\mylen
\newlist{mycases}{enumerate}{1}
\setlist[mycases,1]{label=\textbf{Case~\arabic*.}, 
    labelwidth=\dimexpr-\mylen-\labelsep\relax,leftmargin=0pt,align=right}

\begin{document}
\title{
    Nematic Phase Transitions in 1D Flat Band Condensates
}
\author{Yeongjun Kim\,\orcidlink{0000-0002-3376-4790}}
    \email{yeongjun.kim.04@gmail.com}
    \affiliation{\pcsadd}
    \affiliation{\ctiqsadd}

\author{Oleg I. Utesov\,\orcidlink{0000-0003-0910-5274}}
    \email{utiosov@gmail.com}
    \affiliation{\pcsadd}
    \affiliation{Department of Physics, Korea Advanced Institute of Science and Technology, Daejeon 34141, Republic of Korea}

\author{Alexei~Andreanov\,\orcidlink{0000-0002-3033-0452}}
    \email{aalexei@ibs.re.kr}
    \affiliation{\pcsadd}
    \affiliation{\ctiqsadd}
    \affiliation{\ustadd}

\author{Mikhail V. Fistul\,\orcidlink{0000-0002-0265-2534}}
    \email{Mikhail.Fistoul@rub.de}
    \affiliation{Theoretische Physik III, Ruhr-Universit\"at Bochum, Bochum 44801, Germany}

\author{Sergej Flach\,\orcidlink{0000-0003-1710-3746}}
    \email{sflach@ibs.re.kr}
    \affiliation{\pcsadd}
    \affiliation{\ctiqsadd}
    \affiliation{\ustadd}
    \affiliation{Centre for Theoretical Chemistry and Physics, The New Zealand Institute for Advanced Study (NZIAS), 
    Massey University Albany, Auckland 0745, New Zealand}

\date{\today}

\begin{abstract}
    We investigate the ground-state properties of one-dimensional Gross-Pitaevskii flat-band lattices which are parametrized through their compact localized state (CLS) amplitudes.
    We uncover a CLS geometry-driven phase transition into a macroscopically degenerate nematic state with broken time-reversal symmetry.
    The transition is marked by the appearance of constant-density flat-band states and a vanishing sound velocity.   
    We demonstrate that even infinitesimal onsite interactions can destabilize a $k=0$ plane wave condensate, driving the system into a nematic manifold, which persists for any interaction strength. 
    For the particular choice of constant density CLSs which can tile the lattice, we identify additional families of continuously degenerate ground states characterized by vanishing phase stiffness.
    Utilizing Bogoliubov-de Gennes excitations and parallel tempering, we show that these tiling phases are thermally selected at low temperatures via an order-by-disorder mechanism.
    We exemplify our findings for different classes of flat bands.
    Our findings also reveal that the sound velocity in flat-band condensates is a sensitive probe of the underlying nematic phase transitions. 
\end{abstract}

\maketitle

\sect{Introduction}
Flat bands (FBs) are tight-binding lattices with strictly dispersionless energy bands. 
They result in vanishing group velocity and a divergent effective mass, suppressing transport and diffusion~\cite{derzhko2015strongly,leykam2018artificial,rhim2021singular,danieli2024flat,danieli2026progress}.
For finite-range hopping, FBs support compact localized states (CLSs)~\cite{read2017compactly}, whose amplitudes vanish exactly outside a finite region due to destructive interference.
Flat bands come in three classes imprinted by their CLS set properties: orthogonal, linearly independent (but not orthogonal), and singular~\cite{danieli2024flat,danieli2026progress}.
FBs exhibit a nontrivial rich response to external perturbations, manifesting in novel phases of superconducting, magnetic, topological, disordered, and other matter~\cite{cao2018unconventional,xie2021fractional,derzhko2015strongly,rhim2020quantum,chalker2010anderson,kuno2020flat_qs,danieli2018compact}.
FBs have been implemented in a broad range of different experimental platforms~\cite{nakata2012observation,mukherjee2015observation,vicencio2015observation,taie2015coherent,baboux2016bosonic,wang2022observation,zhou2023observation,lape2025realization}.

In bosonic FBs, interaction effects arise in at least two distinct regimes. 
At low filling, flat bands can stabilize commensurate charge density wave states, whose doping leads to defect-driven physics involving domain walls or interstitial particles~\cite{huber2010bose}. 
In the opposite weak interaction, mean-field regime with macroscopic occupation, or Gross-Pitaevskii (GP) regime, FB condensates are shown to exhibit a finite sound velocity despite the flatness using quantum geometry arguments~\cite{julku2021quantum}.
We show that the CLS/FB geometry has a crucial impact on the appearance of novel degenerate nematic ground states and vanishing sound velocities.


In this manuscript, we study Gross-Pitaevskii condensates on (i) one-dimensional orthogonal all-bands-flat (ABF) lattices, where an angle \(\theta\) continuously tunes the relative amplitudes of the compact localized states and the underlying flat-band geometry, and (ii) on the linearly independent sawtooth lattice~
\footnote{
    Here, the term ``geometry'' refers to the  properties of the CLS vector set.
    Changing the CLS set geometry does not require any changes in the band structure and appears not to be specifically related to quantum geometry.
}.
We show that varying \(\theta\) drives a nontrivial phase transition of the ground state, from a \(k=0\) plane-wave condensate to a nematic phase~\cite{carpenter1989phase,vzukovivc2019xy} with constant density, macroscopic phase degeneracy, and broken time-reversal symmetry.
Additional families of continuously degenerate tiling ground states emerge at special endpoint  \(\theta\) values where the CLS amplitudes become equal, and the CLS (sub)set is tiling the lattice, with free choice of the CLS phases.
The tiling states are characterized by a vanishing sound velocity and are thermally favored in a finite neighborhood of the above endpoint.
Competition between qualitatively distinct sound velocity responses can be used as an experimental way of detection.
Using the sawtooth lattice as an example, we further show that such nontrivial condensate phases exist beyond the ABF lattice architecture and are of generic nature.

\sect{Ground states of 1D GP flat bands}
In the grand canonical setting, the ground state is obtained by minimizing the grand potential \(\mc L(\psi)=\mc H(\psi)-\mu \mc N(\psi)\), where \(\mc H\) is the Hamiltonian, \(\mu\) is the chemical potential, and \(\mc N(\psi)\) is the conserved particle number.  
The stationarity condition is given by \(\partial \mc L/\partial \psi_l^* = 0\).

Let us consider a one-dimensional nearest-neighbor two-band all-bands-flat (ABF) lattice of size \(L\) with periodic boundary conditions~\cite{cadez2021metal, lee2023critical, kim2023flat}.
We assume that the lowest flat band is separated from the rest of the spectrum by a gap.

The sublattice amplitudes are denoted by \(a_l\) and \(b_l\) which are complex numbers, corresponding to expectation values of creation-annihilation operators $a^\dagger_l,a_l, b^\dagger_l, b_l$, and we write \(\psi = (a_1,b_1,a_2,b_2,\ldots,a_L,b_L)\).
Without loss of generality, we set two flat band energies to \(E_f=1\) and \(E_p=0\), respectively, so that the band gap is \(\dgap=E_f-E_p=1\). 
The particle number is \(\mc N(\psi)=\sum_l \left(|a_l|^2+|b_l|^2\right)\).

The quadratic ABF Hamiltonian parametrically depends on an angle $\theta$ and is given by~\cite{cadez2021metal, lee2023critical, kim2023flat}
\begin{align}
    \label{eq:abf_ham}
    \mc H_2(\psi)
    &=
    \sum_l \left[
        \epsilon_a |a_l|^2
        +\epsilon_b |b_l|^2
        +\left(
            t_0 a_l^*b_l
            +t_1 a_l^*a_{l-1}
    \right. \right.
    \notag \\
    &\qquad
    \left.
    \left.
        +t_2 b_l^*b_{l-1}
        +t_3 a_l^*b_{l-1}
        +t_4 b_l^*a_{l-1}
        +\mathrm{h.c.}
    \right)
    \right]
    \notag \\
    &=
    \sum_l
    \left|
        cs\,a_l-s^2b_l+cs\,a_{l-1}+c^2b_{l-1}
    \right|^2,
\end{align}
where \(c=\cos\theta\), \(s=\sin\theta\),
    \(\epsilon_a = 2c^2s^2\), 
    \(\epsilon_b = c^4+s^4\), 
    \(t_0 = cs(c^2-s^2)\), 
    \(t_1 = c^2s^2\), \(t_2 = -c^2s^2\),
    \(t_3 = c^3s\), and \(t_4 = -cs^3\).
The onsite GP interaction is
\begin{equation}
    \label{eq:h4}
    \mc{H}_4(\psi)=\frac{g}{2}\sum_l \left(|a_l|^4+|b_l|^4\right),
\end{equation}
and the total classical Hamiltonian is the sum of both: \(\mc{H}(\psi)=\mc H_2(\psi)+\mc H_4(\psi).\)

We restrict \(\theta\) to the irreducible range \(0 \leq \theta \leq \pi/4\) in~\eqref{eq:abf_ham}, since 
\(\theta \to \theta+\pi\) is a periodic redundancy,
and \(\theta \to \pi/2-\theta\) results in a mirror-symmetric transformation with equivalent physics for our GP problem.

\begin{figure*}
    \centering
    \includegraphics[trim={0.05cm 0.1cm 0.04cm 0.14cm}, clip,width=\linewidth]{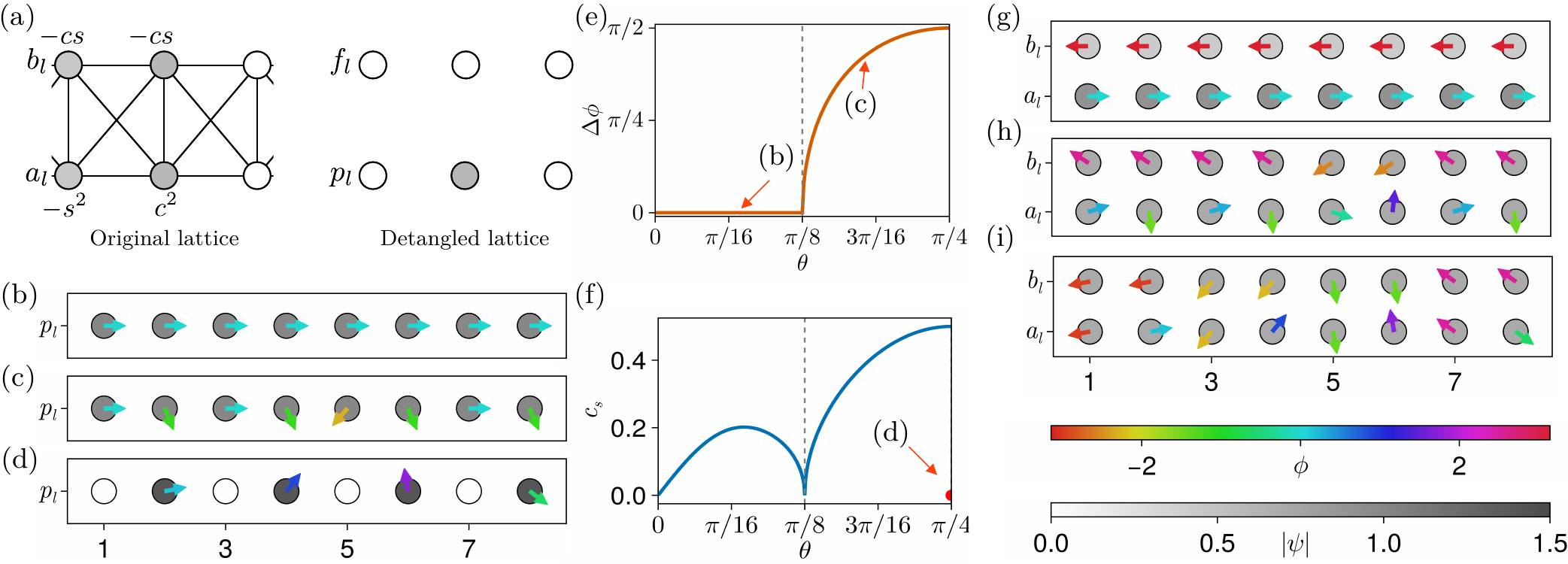}
    \caption{
        Summary of the GP-ABF results.
        (a) ABF tight-binding lattice in the original (left) and detangled (right) bases.
        Shaded regions indicate lower-band CLSs~\eqref{eq:cls}, with the site labels showing their amplitudes.
        Here, \(c=\cos\theta\) and \(s=\sin\theta\).
        (b)--(c) Representative ground states at \(\theta/\pi=0.07\), \(0.16\), as marked in (e), in the projected detangled basis.
        (d) Tiling ground state in the projected basis, at \(\theta/\pi=1/4\).
        (e) Ground-state phase diagram, showing \(\Delta\phi\equiv|\Delta\phi_l|\) as a function of \(\theta\).
        The dashed line marks the transition at \(\theta=\pi/8\).
        (f) Sound velocity \(c_s\) in the plane wave and nematic regimes.
        The red circle denotes the density-modulated state at \(\theta=\pi/4\).
        (g)--(i) The same ground states as in (b)--(d), but now in the original basis.
    }
    \label{fig:abf_summary}
\end{figure*}

The following local unitary real-space transformation diagonalizes (``detangles'')~\cite{flach2014detangling} \(\mc{H}_2\) of the ABF lattice for any \(\theta\)~\cite{cadez2021metal,kim2023flat, lee2023critical, flach2014detangling}:
\begin{align}
    \label{eq:cls}
    \notag p_l &= c^2 a_l - cs\, b_l - s^2 a_{l-1} - cs\, b_{l-1}, \\
    f_l &= cs\, a_l - s^2 b_l + cs\, a_{l-1} + c^2 b_{l-1}.
\end{align}
Here, \(p_l\) and \(f_l\) correspond to the orthogonal CLS of the lower and upper flat bands, respectively.
Conversely, we can ``entangle'' back to the original basis
\begin{align}
    \label{eq:anbn_in_terms_of_pnfn}
    a_l &= c^2 p_l + cs f_l - s^2 p_{l+1} + cs f_{l+1}, \notag \\
    b_l &= -cs p_l - s^2 f_l - cs p_{l+1} + c^2 f_{l+1}.
\end{align}
Under this transformation, \mbox{Eq.~\eqref{eq:abf_ham}} reduces to
\begin{align}
    \mc{H}_2(\psi)=\sum_l E_f |f_l|^2 + E_p|p_l|^2.
\end{align}
The $E_p=0$ FB states \(p_l=\delta_{l,l_0}\) and \(E_f=1\) states \(f_l=\delta_{l,l_0}\) are compact localized states centered at \(l_0\).  
Therefore, any configuration satisfying \(f_l=0\) minimizes \(\mc{H}_2(\psi)\) and belongs to the lowest flat band manifold, see \mbox{Fig.~\ref{fig:abf_summary}(a)} for the lattice structures in the original and detangled bases.

With the average wave function (norm) density \(n=N/L\) (where \(N = \mc N(\psi)\) is the total norm), the weak interaction regime is set by the interaction energy scale \(gn \ll \dgap\).  
Then the ground state is obtained by minimizing the GP interaction term within the flat band subspace \(\mc{H}_2(\psi)=0\).  
Thus,
\begin{align}
    \label{eq:gs_small_g}
    G = \argmin_{\mc{H}_2(\psi)=0}\bigg[\mc{H}_4(\psi)-\mu \mc{N}(\psi)\bigg].
\end{align}
Equivalently, denoting by \(\leff\), \(\heff\), and \(\neff\) the projected energy functionals on the flat band manifold, we minimize
\begin{align}
    \leff = \heff(\tilde{\psi}) - \mu \neff(\tilde{\psi}),
\end{align}
where, in this case, \(\tilde{\psi}=(p_1,p_2,\ldots)\) is the projected wave function in the CLS basis.  
In this projected description, the nonlinearity strength \(g\) enters only as an overall energy scale.
Therefore 
\begin{align}
    \label{eq:abf1d_gp_projected}
    \leff
    = \sum_l \frac{g}{2}\big(|c^2 p_l - s^2 p_{l+1}|^4 + (cs)^4 |p_l + p_{l+1}|^4\big) - \mu |p_l|^2.
\end{align}
Note that the norm density \(\neff\) is independent of relative phases of \(p_l\) and \(p_{l+1}\) in ABF.
By expanding the quartic terms, one finds that the projected interaction induces effective nearest-neighbor couplings $\tilde{t} p_l p_{l+1}^*$ between the CLS amplitudes.  
Such geometry-induced couplings are the origin of many nontrivial flat band phenomena in perturbed settings, including flat band superconductivity and disorder-driven delocalization~\cite{julku2021quantum, peotta2015superfluidity, cadez2021metal}.
In our case, the effective hoppings $\tilde{t}$ are products of the CLS amplitudes and fall into two distinct classes: density-assisted hoppings which do not depend on the CLS phases, and phase-assisted hoppings which do depend on the CLS phases. 
The resulting competition between these different hoppings is unaffected by the smallness of the interaction $g$ and is the physical origin of the nematic phase transition (see SM~\cite{supp_local} for an illustrative mapping onto the XY spin model).

\(\leff\)~\eqref{eq:abf1d_gp_projected} is minimized by a state with uniform norm density (using the Cauchy-Schwarz inequality, see SM~\cite{supp_local})
\begin{align}
    \label{eq:uniform_norm_density}
    p_l = \sqrt{n}\,e^{i\phi_l},
\end{align}
with \(\phi_{l+1}=\phi_l+\Delta\phi_l\). 
The optimal phase difference is 
\begin{align}
    \label{eq:dphi}
    \Delta \phi_l =
    \begin{cases}
        0, & 0 \leq \theta \leq \frac{\pi}{8}, \\
        \sigma_l\arccos\big(\cot^2(2\theta)\big) & \frac{\pi}{8} \leq \theta < \frac{\pi}{4}.
    \end{cases}
\end{align}
with \(\sigma_l \in\{-1, 1\}\) [\mbox{Fig.~\ref{fig:abf_summary}}(e)].
Thus, for \(0 \leq \theta \leq \pi/8\), the ground state of \eqref{eq:abf1d_gp_projected} is a $k=0$ plane wave condensate in the detangled projected space [\mbox{Fig.~\ref{fig:abf_summary}}(b)].
In the original lattice, this ground state manifests as one nondegenerate $k=0$ state with different densities and opposite phases on the two legs [\mbox{Fig.~\ref{fig:abf_summary}}(g)].
For \(\pi/8 \leq \theta < \pi/4\), the ground state forms a macroscopically degenerate nematic manifold with random sequences of nonzero local phase differences.
Each condensate has constant norm density both in the projected space  \mbox{Fig.~\ref{fig:abf_summary}}(c) and in the original lattice \mbox{Fig.~\ref{fig:abf_summary}}(h).
The chemical potential is given by~\cite{supp_local}
\begin{align}
    \label{eq:gsenergy_chemical_potential}
    \mu &= \frac{gn}{2}
    \begin{cases}
        2\big(\cos^4 (2\theta) + \sin^4(2\theta)\big), & 0 \leq \theta \leq \frac{\pi}{8}, \\
        1, & \frac{\pi}{8} \leq \theta \leq \frac{\pi}{4}.
    \end{cases}
\end{align}
The presence of nonzero phase differences \(\Delta\phi_l\) in the projected model translates into local currents in the original flat band network.
This makes the nematic phase a broken time-reversal state with local synthetic fluxes~\cite{supp_local}.

For even \(L\) and periodic boundary conditions, the nematic ground state degeneracy holds if \(\sum_l \Delta\phi_l = 2\pi m\), where \(m\) is an integer winding number.
For $m=0$ the degeneracy multiplicity is \(C^L_{L/2}\) (here, \(C\) denotes combination). 
For $m\neq 0$ only rational sets of $\Delta\phi_l$ survive. 
For odd \(L\), the residual mismatch causes a slight increase of order \(\text{O}(1/L)\) in the ground state energy. 

The nematic ground states ~\eqref{eq:uniform_norm_density}-\eqref{eq:dphi} are exact ground states for any strength of the interaction $g>0$. 
This surprising fact follows from the CLS and FB geometry and applies to a plethora of FB models.
All it needs is a homogeneous FB eigenstate with constant norm density profile in the original lattice basis, since such a state minimises simultaneously both $\mc H_2$ and $\mc H_4$~\cite{supp_local}.
A FB model manifold can be parametrized by its own CLS~\cite{maimaiti2017compact,maimaiti2019universal,maimaiti2021flatband,graf2021designing,hwang2021general}.
Here a CLS is given by its four amplitudes $A_1,B_1,A_2,B_2$.
A homogeneous FB state is obtained from a linear combination of CLSs if $|A_2+A_1 \mathrm{e}^{i\phi}|^2 = |B_2+B_1\mathrm{e}^{i\phi}|^2$.
For real valued $A_{1,2},B_{1,2}$ it follows \(\cos \phi=(A_1^2+A_2^2-B_1^2-B_2^2)/2(B_1B_2-A_1A_2)\). 
For a sufficiently inhomogeneous CLS profile, no solution for $\phi$ exists~\cite{supp_local}. 
Tuning the CLS to a more homogeneous one, a transition emerges at $(A_1\pm A_2)^2= (B_1\pm B_2)^2$ with $\phi = 0,\pi$.
Even more homogeneous CLS choices result in two $\pm \phi$ solutions and a nematic phase scenario.
The $\phi=0$ transition is akin to our above nematic phase transition, while the $\phi=\pi$ transition is yet to be observed. 

We verified the analytical solutions in \mbox{Eqs.~\eqref{eq:dphi}} and \eqref{eq:gsenergy_chemical_potential} numerically. 
To obtain a representative family of solutions, we prepare low-temperature samples based on Langevin Monte Carlo~\cite{xu2020advancedhmc} with random initial conditions combined with parallel tempering~\cite{earl2005parallel, hukushima1996exchange}, and further minimize by gradient descent-type optimization.
Examples of nematic-phase configurations are shown in Fig.~\ref{fig:abf_summary}(c) for \(\theta/\pi= 0.16\).
Using \mbox{Eq.~\eqref{eq:anbn_in_terms_of_pnfn}} with \(f_l = 0\), we can also obtain the corresponding condensate profile in the original \((a_l,b_l)\) basis, see SM~\cite{supp_local}.

Next we estimate the energy barriers between possible nematic ground states.
We perform local phase variations and estimate the saddle point energy when passing from a given ground state to its local neighbour by continuously varying a local $\sigma_l$ into its negative.
The saddle point is achieved  by zeroing the phase difference on this bond, resulting in
\begin{gather}
    \Delta E = \frac{g n^2}{4} \left[ 2(1- 2 \sin^2{2\theta} +2 \sin^4{2\theta}) - 1 \right].
\end{gather}
The energy barrier grows starting from $0$ at $\theta = \pi/8$ to $g n^2/4$ for $\theta = \pi/4$.
Our estimate neglects density modulations, which are expected to be costly in general, but may start to compete with phase variations close to the endpoint $\theta=\pi/4$.

At \(\theta=\pi/4\), an additional pair of ground state families appears beyond the nematic configurations discussed above. 
At this point, the CLS amplitudes \(p_l\) of the lower-energy flat band [\mbox{Eq.~\eqref{eq:cls}}] become homogeneous, with equal magnitudes on all four sites. 
One can therefore tile the entire original FB lattice with non-overlapping CLSs on alternating unit cells, resulting in a density modulation in the projected space [Fig.~\ref{fig:abf_summary}(d)] while maintaining uniform norm densities in the original \((a_l,b_l)\) basis [Fig.~\ref{fig:abf_summary}(i)]. 
Being homogeneous, these states again minimize \(\mathcal H_4\), and persist as exact ground states for arbitrary interaction strength $g$. 
This endpoint therefore hosts an enlarged ground state manifold containing, in addition to the degenerate nematic state set, two families of \emph{tiling ground states} with $L/2$ continuous CLS phase degeneracies. 
Close to the endpoint \(\theta=\pi/4\) these states form a platform for a high-dimensional manifold of low-energy excitations which dominate the low-temperature statistics as shown below.

\sect{BdG excitations}
We now examine how the condensate transition discussed above is reflected in the Bogoliubov-de Gennes (BdG) excitations, which govern the low-energy dynamics around the ground states.

The low-energy description of the BdG excitations (Goldstone modes) is governed by two hydrodynamic parameters, the phase stiffness \(\rho_s\) and the compressibility \(\kappa\)~\cite{supp_local}. 
Here \(\rho_s\) characterizes the phase rigidity of the condensate, while \(\kappa\) determines its density response. 
To characterize the phase rigidity of the condensate, we define the phase stiffness \(\rho_s\) from the energy cost of a long-wavelength phase twist, \(\delta E = \frac{1}{2}\rho_s \int dx\, (\nabla \phi)^2\),
or equivalently from the second derivative of the ground-state energy with respect to a uniform twist~\cite{supp_local}. 
For the stiffness, we obtain
\begin{align}
    \label{eq:stiffness}
    \rho_s = \frac{gn^2}{2}
    \begin{cases}
        \cos(4\theta)\sin^2(2\theta), & 0 \le \theta < \frac{\pi}{8}, \\
        |\cos(4\theta)|, & \frac{\pi}{8} \le \theta < \frac{\pi}{4}.
    \end{cases}
\end{align}
In the continuum limit, the sound velocity is related to the stiffness and compressibility through \(c_s^2 = \rho_s/\kappa\),
where \(\kappa = \partial n/\partial \mu\) is the compressibility~\cite{geier2025superfluidity,julku2021excitations}.

We also note that \(\rho_s\) can be re-expressed as a function of quantum metric ~\cite{julku2021quantum,peotta2015superfluidity}; for completeness, we record the corresponding formula in the SM~\cite{supp_local}. 

At the special point \(\theta=\pi/4\), in addition to the nematic states, the tiling ground state manifold emerges. 
This manifold has vanishing stiffness, \(\rho_s = 0\), because in the \(p_l\) basis the condensates are isolated on every second site (unit cell), so a phase twist costs no energy.

We verify these analytical results by numerically computing the BdG spectrum. 
The spectrum is obtained by linearizing the dynamics around a ground state \(G_l\) and using the decomposition
\begin{align}
    \label{eq:GS_fluctuation_spectrum}
    \delta G_l(t)=\chi_l e^{-i\lambda t}+\Pi_l^* e^{i\lambda t}.
\end{align}
Because of global phase invariance, the BdG spectrum is always gapless. 
The sound velocity \(c_s\) is defined by the slope of the Goldstone mode at low momentum, \(\lambda(k)\approx c_s |k|\). 
The full derivation of the BdG equations is provided in the End Matter.

Figure~\ref{fig:abf_summary}(f) shows the sound velocity \(c_s\) as a function of \(\theta\).
In the homogeneous regime \(0 \le \theta < \pi/8\), \(c_s\) starts from zero at the trivial point \(\theta=0\), then increases, and, quite remarkably, decreases again as \(\theta\) approaches the critical point.
It eventually vanishes at \(\theta=\pi/8\), signaling the instability of the condensate.
In the nematic regime \(\pi/8 < \theta < \pi/4\), \(c_s\) gradually increases, reaching its maximum value at $\theta=\pi/4$. 
At \(\theta=\pi/4\), the additional tiling states have \(c_s=0\) as discussed above (drawn as a red circle).
The sound velocity shown in \mbox{Fig.~\ref{fig:abf_summary}}(f) is compared with the full BdG spectra obtained from exact diagonalization, and we find faithful agreement in the low-energy region, as shown in the End Matter.

We emphasize two notable features of this result.
At \(\theta=\pi/8\), the entire BdG spectrum collapses to zero, signaling the instability associated with the phase transition.
Moreover, for fixed \(\theta\) in the nematic regime, the sound velocity \(c_s\) is identical for all nematic ground states despite their macroscopic degeneracy.

\sect{Low temperature statistics}
The macroscopic degeneracy of the nematic GS manifold naturally raises the following question:
how is this manifold statistically weighted at finite temperature $T$?
Do thermal fluctuations favor some nematic configurations over others, perhaps by an order-by-disorder (ObD) mechanism~\cite{villain1980order}? 
The low-energy BdG spectrum and the sound velocity are tools to address this question, since they determine the leading order fluctuation contribution to the free energy.

We study finite-\(T\) statistics of the projected Hamiltonian \(\heff\) by sampling the projected wave function \(\tilde{\psi}\) with Boltzmann weights \(P(\tilde{\psi}) = e^{-\beta \heff(\tilde{\psi})}/\mc Z\), where \( \beta=1/T\) and \(\mc Z\) is the partition function.
At low density and low \(T\) (\(\beta\dgap \gg 1 \)), the particle number \(\mc \neff(\tilde{\psi})\) enters only as an overall scale factor in \(\heff\),
so its fluctuations merely rescale the total energy without affecting the structure of the state space. 
We therefore work in the canonical ensemble with fixed norm to isolate the nontrivial statistical fluctuations.

At \(\theta = \pi/4\), the alternating-density solution is selected by the ObD mechanism because it yields a lower free energy correction to the ground-state energy: 
the densities on alternating sites are effectively decoupled, and the corresponding BdG spectrum becomes completely flat, minimizing the free energy correction. 
Denote the correction by \(\delta F_\mathrm{Th}\), the quantum acoustic-mode contribution of coupled oscillators in one dimension behaves as
\begin{align}
    \label{eq:thermal_correction}
    \delta F_{\mathrm{Th}} \propto -\frac{T^2}{c_s},
\end{align}
as shown in the SM~\cite{supp_local}, so that softer Goldstone modes lead to a lower free energy. 
As discussed above, at \(\theta=\pi/8\) and \(\theta=\pi/4\) there exist states with \(c_s=0\).
This indicates that the linear BdG description is no longer sufficient at these special points (see End Matter for numerical confirmation).

\begin{figure}[t]
    \centering
    \includegraphics[trim={0.1cm 0.2cm 0.12cm 0.1cm}, clip, width=0.99\linewidth]{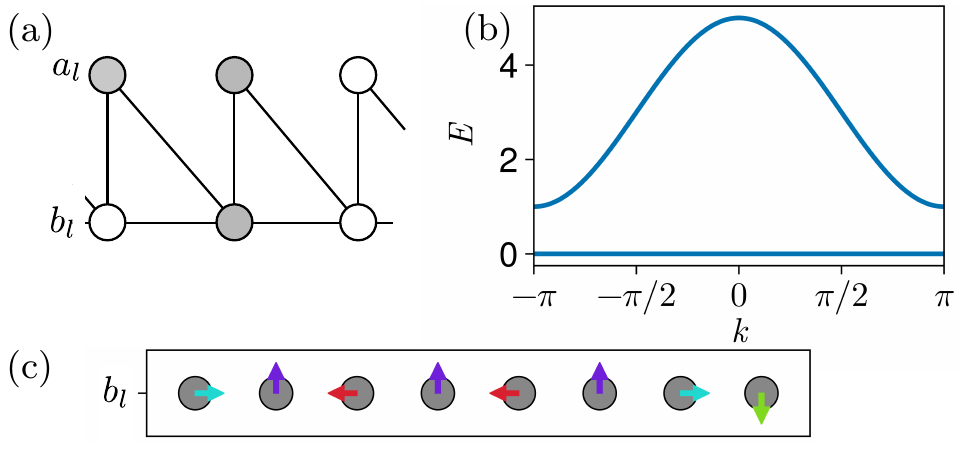}
    \caption{
        (a) Sawtooth chain tight-binding lattice. Black circles denote CLS.
        (b) The band structure of the sawtooth chain.
        (c) Numerically obtained ground state of the GP-sawtooth chain, only showing \(b\) sublattices.
        The arrows represent complex amplitudes in the complex plane.
        The arrow angle color code is the same as in Fig.~\ref{fig:abf_summary}.
    }
    \label{fig:sawtooth}
\end{figure}

\sect{Sawtooth chain}
We now show that the nematic ground state structure persists for linearly independent flat band networks with a mixed band structure, and is therefore not a privilege of orthogonal ABFs.
We consider the linearly independent sawtooth chain~\cite{huber2010bose} as a representative example showing one flat and one dispersive band in the non-interacting regime (\(g=0\)) \mbox{Fig.~\ref{fig:sawtooth}}(a,b).
The Hamiltonian of the GP-sawtooth lattice is given by
\begin{equation}
    \label{eq:sawtooth_gp_sum_of_square}
    \mc H = \frac{1}{2}\sum_l \left|\sqrt{2}\,a_l + b_l + b_{l+1}\right|^2 + g\sum_l \left(|a_l|^4 + |b_l|^4\right).
\end{equation}
As we outline in the SM~\cite{supp_local}, the model hosts a degenerate set of nematic ground states with uniform norm density.
The wave function is again highly degenerate, with the phase difference \(\Delta\phi=\pm\pi/2\) on the \(b\)-sublattice.
For the \(a\)-sublattice, the flat band constraint yields \(a_l = -\sqrt{n_b}\,e^{i\phi_l+\Delta\phi_{l+1}/2}\) with \(n_b=\mu/(2g)\). 
Thus, the ground state of the GP-sawtooth lattice choice corresponds to a particular nematic phase case, see Fig.~\ref{fig:sawtooth}(c).
Modifying the CLS amplitudes results in a family of generalized sawtooth FBs~\cite{maimaiti2017compact} which host a nematic phase transition at which the CLS profiles become sufficiently inhomogeneous and the FB ceases to host homogeneous density states similar to the above ABF example.

\sect{Conclusions}
We studied the GP problem of one-dimensional FB lattices and showed that varying the flat band geometry/CLS parameter \(\theta\) gives rise to multiple competing condensate phases, including $k=0$ plane waves, time-reversal-broken nematic and tiling states. 
We further found that these phases exhibit distinct excitation spectra and sound velocity responses. 
Our results show that the sound velocity \(c_s\) of flat band condensates depends sensitively on the GS phase structure. 
At finite temperatures and close to transition points it is governed by order-by-disorder mechanisms.
Our results may provide a useful perspective in the broader field of flat band superconductivity, where superfluid transport can remain nontrivial despite the quenched kinetic energy. 

While preparing this manuscript, we became aware of a recent work that discusses the nematic BEC phase in two-dimensional flat bands~\cite{huhtinen2026stability2}. 

\begin{acknowledgments}
    The authors acknowledge financial support from the Institute for Basic Science (IBS) in the Republic of Korea through Project No. IBS-R024-D1 and IBS-R041-D1-2026-a00.
    M.V.F. thanks the hospitality of the IBS Center for Theoretical Physics of Complex Systems. 
\end{acknowledgments}

\bibliographystyle{apsrev4-2}
\bibliography{local}

@article{derzhko2015strongly,
  author =        {Derzhko, Oleg and Richter, Johannes and
                   Maksymenko, Mykola},
  journal =       {Int. J. Mod. Phys. B},
  number =        {12},
  pages =         {1530007},
  title =         {Strongly correlated flat-band systems: The route from
                   Heisenberg spins to Hubbard electrons},
  volume =        {29},
  year =          {2015},
  doi =           {10.1142/S0217979215300078},
  url =           {http://www.worldscientific.com/doi/abs/10.1142/
                  S0217979215300078},
}

@article{leykam2018artificial,
  author =        {Leykam, Daniel and Andreanov, Alexei and
                   Flach, Sergej},
  journal =       {Adv. Phys.: X},
  number =        {1},
  pages =         {1473052},
  publisher =     {Taylor \& Francis},
  title =         {Artificial flat band systems: from lattice models to
                   experiments},
  volume =        {3},
  year =          {2018},
  doi =           {10.1080/23746149.2018.1473052},
  url =           {https://doi.org/10.1080/23746149.2018.1473052},
}

@article{rhim2021singular,
  author =        {Rhim, Jun-Won and Yang, Bohm-Jung},
  journal =       {Advances in Physics: X},
  number =        {1},
  pages =         {1901606},
  publisher =     {Taylor \& Francis},
  title =         {Singular flat bands},
  volume =        {6},
  year =          {2021},
  doi =           {10.1080/23746149.2021.1901606},
  url =           {https://doi.org/10.1080/23746149.2021.1901606},
}

@article{danieli2024flat,
  author =        {Carlo Danieli and Alexei Andreanov and Daniel Leykam and
                   Sergej Flach},
  journal =       {Nanophotonics},
  number =        {21},
  pages =         {3925--3944},
  title =         {Flat band fine-tuning and its photonic applications},
  volume =        {13},
  year =          {2024},
  doi =           {doi:10.1515/nanoph-2024-0135},
  url =           {https://doi.org/10.1515/nanoph-2024-0135},
}

@misc{danieli2026progress,
  author =        {Carlo Danieli and Sergej Flach},
  title =         {Progress on artificial flat band systems:
                   classifying, perturbing, applying},
  year =          {2026},
  url =           {https://arxiv.org/abs/2603.04248},
}

@article{read2017compactly,
  author =        {Read, N.},
  journal =       {Phys. Rev. B},
  month =         mar,
  pages =         {115309},
  publisher =     {American Physical Society},
  title =         {Compactly supported Wannier functions and algebraic
                   $K$-theory},
  volume =        {95},
  year =          {2017},
  doi =           {10.1103/PhysRevB.95.115309},
  url =           {https://link.aps.org/doi/10.1103/PhysRevB.95.115309},
}

@article{cao2018unconventional,
  author =        {Cao, Yuan and Fatemi, Valla and Fang, Shiang and
                   Watanabe, Kenji and Taniguchi, Takashi and
                   Kaxiras, Efthimios and Jarillo-Herrero, Pablo},
  journal =       {Nature},
  pages =         {43--50},
  title =         {Unconventional superconductivity in magic-angle
                   graphene superlattices},
  volume =        {556},
  year =          {2018},
  doi =           {10.1038/nature26160},
  url =           {https://doi.org/10.1038/nature26160},
}

@article{xie2021fractional,
  author =        {Xie, Yonglong and Pierce, Andrew T and
                   Park, Jeong Min and Parker, Daniel E and
                   Khalaf, Eslam and Ledwith, Patrick and Cao, Yuan and
                   Lee, Seung Hwan and Chen, Shaowen and
                   Forrester, Patrick R and others},
  journal =       {Nature},
  number =        {7889},
  pages =         {439--443},
  publisher =     {Nature Publishing Group UK London},
  title =         {Fractional Chern insulators in magic-angle twisted
                   bilayer graphene},
  volume =        {600},
  year =          {2021},
}

@article{rhim2020quantum,
  author =        {Rhim, Jun-Won and Kim, Kyoo and Yang, Bohm-Jung},
  journal =       {Nature},
  month =         aug,
  pages =         {59 -- 63},
  title =         {Quantum distance and anomalous {Landau} levels of
                   flat bands},
  volume =        {584},
  year =          {2020},
  doi =           {10.1038/s41586-020-2540-1},
  url =           {https://doi.org/10.1038/s41586-020-2540-1},
}

@article{chalker2010anderson,
  author =        {Chalker, J. T. and Pickles, T. S. and Shukla, Pragya},
  journal =       {Phys. Rev. B},
  month =         sep,
  pages =         {104209},
  publisher =     {American Physical Society},
  title =         {Anderson localization in tight-binding models with
                   flat bands},
  volume =        {82},
  year =          {2010},
  doi =           {10.1103/PhysRevB.82.104209},
  url =           {https://link.aps.org/doi/10.1103/PhysRevB.82.104209},
}

@article{kuno2020flat_qs,
  author =        {Kuno, Yoshihito and Mizoguchi, Tomonari and
                   Hatsugai, Yasuhiro},
  journal =       {Phys. Rev. B},
  month =         dec,
  pages =         {241115},
  publisher =     {American Physical Society},
  title =         {Flat band quantum scar},
  volume =        {102},
  year =          {2020},
  doi =           {10.1103/PhysRevB.102.241115},
  url =           {https://link.aps.org/doi/10.1103/PhysRevB.102.241115},
}

@article{danieli2018compact,
  author =        {Danieli, C. and Maluckov, A. and Flach, S.},
  journal =       {Low Temp. Phys.},
  number =        {7},
  pages =         {678-687},
  title =         {Compact discrete breathers on flat-band networks},
  volume =        {44},
  year =          {2018},
  doi =           {10.1063/1.5041434},
}

@article{nakata2012observation,
  author =        {Nakata, Yosuke and Okada, Takanori and
                   Nakanishi, Toshihiro and Kitano, Masao},
  journal =       {Phys. Rev. B},
  month =         may,
  pages =         {205128},
  publisher =     {American Physical Society},
  title =         {Observation of flat band for terahertz spoof plasmons
                   in a metallic kagom\'e lattice},
  volume =        {85},
  year =          {2012},
  doi =           {10.1103/PhysRevB.85.205128},
  url =           {https://link.aps.org/doi/10.1103/PhysRevB.85.205128},
}

@article{mukherjee2015observation,
  author =        {Mukherjee, Sebabrata and Spracklen, Alexander and
                   Choudhury, Debaditya and Goldman, Nathan and
                   \"Ohberg, Patrik and Andersson, Erika and
                   Thomson, Robert R.},
  journal =       {Phys. Rev. Lett.},
  month =         jun,
  pages =         {245504},
  publisher =     {American Physical Society},
  title =         {Observation of a Localized Flat-Band State in a
                   Photonic {Lieb} Lattice},
  volume =        {114},
  year =          {2015},
  doi =           {10.1103/PhysRevLett.114.245504},
  url =           {http://link.aps.org/doi/10.1103/PhysRevLett.114.245504},
}

@article{vicencio2015observation,
  author =        {Vicencio, Rodrigo A. and Cantillano, Camilo and
                   Morales-Inostroza, Luis and Real, Basti\'an and
                   Mej\'{\i}a-Cort\'es, Cristian and Weimann, Steffen and
                   Szameit, Alexander and Molina, Mario I.},
  journal =       {Phys. Rev. Lett.},
  month =         jun,
  pages =         {245503},
  publisher =     {American Physical Society},
  title =         {Observation of Localized States in {Lieb} Photonic
                   Lattices},
  volume =        {114},
  year =          {2015},
  doi =           {10.1103/PhysRevLett.114.245503},
  url =           {http://link.aps.org/doi/10.1103/PhysRevLett.114.245503},
}

@article{taie2015coherent,
  author =        {Shintaro Taie and Hideki Ozawa and Tomohiro Ichinose and
                   Takuei Nishio and Shuta Nakajima and
                   Yoshiro Takahashi},
  journal =       {Science Advances},
  number =        {10},
  pages =         {e1500854},
  title =         {Coherent driving and freezing of bosonic matter wave
                   in an optical Lieb lattice},
  volume =        {1},
  year =          {2015},
  doi =           {10.1126/sciadv.1500854},
  url =           {https://www.science.org/doi/abs/10.1126/sciadv.1500854},
}

@article{baboux2016bosonic,
  author =        {Baboux, F. and Ge, L. and Jacqmin, T. and Biondi, M. and
                   Galopin, E. and Lema\^{\i}tre, A. and Le Gratiet, L. and
                   Sagnes, I. and Schmidt, S. and T\"ureci, H.E. and
                   Amo, A. and Bloch, J.},
  journal =       {Phys. Rev. Lett.},
  month =         feb,
  pages =         {066402},
  publisher =     {American Physical Society},
  title =         {Bosonic Condensation and Disorder-Induced
                   Localization in a Flat Band},
  volume =        {116},
  year =          {2016},
  doi =           {10.1103/PhysRevLett.116.066402},
  url =           {http://link.aps.org/doi/10.1103/PhysRevLett.116.066402},
}

@article{wang2022observation,
  author =        {Wang, Haiteng and Zhang, Weixuan and Sun, Houjun and
                   Zhang, Xiangdong},
  journal =       {Phys. Rev. B},
  month =         {Sep},
  pages =         {104203},
  publisher =     {American Physical Society},
  title =         {Observation of inverse Anderson transitions in
                   Aharonov-Bohm topolectrical circuits},
  volume =        {106},
  year =          {2022},
  doi =           {10.1103/PhysRevB.106.104203},
  url =           {https://link.aps.org/doi/10.1103/PhysRevB.106.104203},
}

@article{zhou2023observation,
  author =        {Zhou, Xiaoqi and Zhang, Weixuan and Sun, Houjun and
                   Zhang, Xiangdong},
  journal =       {Phys. Rev. B},
  month =         {Jan},
  pages =         {035152},
  publisher =     {American Physical Society},
  title =         {Observation of flat-band localization and topological
                   edge states induced by effective strong interactions
                   in electrical circuit networks},
  volume =        {107},
  year =          {2023},
  doi =           {10.1103/PhysRevB.107.035152},
  url =           {https://link.aps.org/doi/10.1103/PhysRevB.107.035152},
}

@article{lape2025realization,
  author =        {Lape, Noah and Diubenkov, Simon and English, L. Q. and
                   Kevrekidis, P. G. and Andreanov, Alexei and
                   Kim, Yeongjun and Flach, Sergej},
  journal =       {Phys. Rev. B},
  month =         {Nov},
  pages =         {184309},
  publisher =     {American Physical Society},
  title =         {Realization and characterization of an all-bands-flat
                   electrical lattice},
  volume =        {112},
  year =          {2025},
  doi =           {10.1103/1w5c-nsmh},
  url =           {https://link.aps.org/doi/10.1103/1w5c-nsmh},
}

@article{huber2010bose,
  author =        {Huber, Sebastian D. and Altman, Ehud},
  journal =       {Phys. Rev. B},
  month =         nov,
  pages =         {184502},
  publisher =     {American Physical Society},
  title =         {Bose condensation in flat bands},
  volume =        {82},
  year =          {2010},
  doi =           {10.1103/PhysRevB.82.184502},
  url =           {https://link.aps.org/doi/10.1103/PhysRevB.82.184502},
}

@article{julku2021quantum,
  author =        {Julku, Aleksi and Bruun, Georg M. and
                   T\"orm\"a, P\"aivi},
  journal =       {Phys. Rev. Lett.},
  month =         {Oct},
  pages =         {170404},
  publisher =     {American Physical Society},
  title =         {Quantum Geometry and Flat Band Bose-Einstein
                   Condensation},
  volume =        {127},
  year =          {2021},
  doi =           {10.1103/PhysRevLett.127.170404},
  url =           {https://link.aps.org/doi/10.1103/PhysRevLett.127.170404},
}

@article{cadez2021metal,
  author =        {{\v C}ade{\v z}, Tilen and Kim, Yeongjun and
                   Andreanov, Alexei and Flach, Sergej},
  journal =       {Phys. Rev. B},
  month =         {Nov},
  pages =         {L180201},
  publisher =     {American Physical Society},
  title =         {Metal-insulator transition in infinitesimally weakly
                   disordered flat bands},
  volume =        {104},
  year =          {2021},
  doi =           {10.1103/PhysRevB.104.L180201},
  url =           {https://link.aps.org/doi/10.1103/PhysRevB.104.L180201},
}

@article{lee2023critical,
  author =        {Lee, Sanghoon and Andreanov, Alexei and
                   Flach, Sergej},
  journal =       {Phys. Rev. B},
  month =         {Jan},
  pages =         {014204},
  publisher =     {American Physical Society},
  title =         {Critical-to-insulator transitions and fractality
                   edges in perturbed flat bands},
  volume =        {107},
  year =          {2023},
  doi =           {10.1103/PhysRevB.107.014204},
  url =           {https://link.aps.org/doi/10.1103/PhysRevB.107.014204},
}

@article{kim2023flat,
  author =        {Kim, Yeongjun and
                     \ifmmode \check{C}\else \v{C}\fi{}ade\ifmmode
  \check{z}\else \v{z}\fi{}, Tilen and Andreanov, Alexei and Flach, Sergej},
  journal =       {Phys. Rev. B},
  month =         {May},
  pages =         {174202},
  publisher =     {American Physical Society},
  title =         {Flat band induced metal-insulator transitions for
                   weak magnetic flux and spin-orbit disorder},
  volume =        {107},
  year =          {2023},
  doi =           {10.1103/PhysRevB.107.174202},
  url =           {https://link.aps.org/doi/10.1103/PhysRevB.107.174202},
}

@article{flach2014detangling,
  author =        {Flach, Sergej and Leykam, Daniel and
                   Bodyfelt, Joshua D. and Matthies, Peter and
                   Desyatnikov, Anton S.},
  journal =       {Europhys. Lett.},
  number =        {3},
  pages =         {30001},
  title =         {Detangling flat bands into {Fano} lattices},
  volume =        {105},
  year =          {2014},
  url =           {http://stacks.iop.org/0295-5075/105/i=3/a=30001},
}

@article{peotta2015superfluidity,
  author =        {Peotta, Sebastiano and
                   T{\"{o}}rm{\"{a}}, P{\"{a}}ivi},
  journal =       {Nat. Comm.},
  month =         nov,
  pages =         {8944},
  publisher =     {Nature Publishing},
  title =         {Superfluidity in topologically nontrivial flat bands},
  volume =        {6},
  year =          {2015},
  url =           {https://doi.org/10.1038/ncomms9944 http://10.0.4.14/
                  ncomms9944 http://www.nature.com/articles/
                  ncomms9944{\#}supplementary-information},
}

@footnote{supp_local,
  note =         {See Supplemental Material at URL to be defined.},
}

@inproceedings{xu2020advancedhmc,
  author =        {Xu, Kai and Ge, Hong and Tebbutt, Will and
                   Tarek, Mohamed and Trapp, Martin and
                   Ghahramani, Zoubin},
  booktitle =     {Symposium on Advances in Approximate Bayesian
                   Inference},
  organization =  {PMLR},
  pages =         {1--10},
  title =         {AdvancedHMC. jl: A robust, modular and efficient
                   implementation of advanced HMC algorithms},
  year =          {2020},
}

@article{earl2005parallel,
  author =        {Earl, David J and Deem, Michael W},
  journal =       {Physical Chemistry Chemical Physics},
  number =        {23},
  pages =         {3910--3916},
  publisher =     {Royal Society of Chemistry},
  title =         {Parallel tempering: Theory, applications, and new
                   perspectives},
  volume =        {7},
  year =          {2005},
}

@article{hukushima1996exchange,
  author =        {Hukushima, Koji and Nemoto, Koji},
  journal =       {J. Phys. Soc. Jpn.},
  number =        {6},
  pages =         {1604-1608},
  publisher =     {The Physical Society of Japan},
  title =         {Exchange Monte Carlo Method and Application to Spin
                   Glass Simulations},
  volume =        {65},
  year =          {1996},
  doi =           {10.1143/JPSJ.65.1604},
}

@article{geier2025superfluidity,
  author =        {Geier, Kevin T and Maki, Jeff and Biella, Alberto and
                   Dalfovo, Franco and Giorgini, Stefano and
                   Stringari, Sandro},
  journal =       {Physical Review Research},
  number =        {1},
  pages =         {013187},
  publisher =     {APS},
  title =         {Superfluidity and sound propagation in disordered
                   Bose gases},
  volume =        {7},
  year =          {2025},
}

@article{julku2021excitations,
  author =        {Julku, Aleksi and Bruun, Georg M. and
                   T\"orm\"a, P\"aivi},
  journal =       {Phys. Rev. B},
  month =         {Oct},
  pages =         {144507},
  publisher =     {American Physical Society},
  title =         {Excitations of a {Bose-Einstein} condensate and the
                   quantum geometry of a flat band},
  volume =        {104},
  year =          {2021},
  doi =           {10.1103/PhysRevB.104.144507},
  url =           {https://link.aps.org/doi/10.1103/PhysRevB.104.144507},
}

@article{villain1980order,
  author =        {Villain, J and Bidaux, R and Carton, J-P and
                   Conte, R},
  journal =       {J. de Phys.},
  number =        {11},
  pages =         {1263--1272},
  publisher =     {Soci{\'e}t{\'e} Fran{{\c c}}aise de Physique},
  title =         {Order as an effect of disorder},
  volume =        {41},
  year =          {1980},
  doi =           {http://dx.doi.org/10.1051/jphys:0198000410110126300},
}

@article{maimaiti2019universal,
 author =         {Maimaiti, Wulayimu and Flach, Sergej and Andreanov, Alexei},
 title =          {Universal $d=1$ flat band generator from compact localized states},
 journal =        {Phys. Rev. B},
 volume =         {99},
 issue =          {12},
 pages =          {125129},
 numpages =       {15},
 year =           {2019},
 month =          {Mar},
 publisher =      {American Physical Society},
 doi =            {10.1103/PhysRevB.99.125129},
 url =            {https://link.aps.org/doi/10.1103/PhysRevB.99.125129},
}

@article{carpenter1989phase,
  title={The phase diagram of a generalised XY model},
  author={Carpenter, DB and Chalker, JT},
  journal={Journal of Physics: Condensed Matter},
  volume={1},
  number={30},
  pages={4907--4912},
  year={1989}
}

@article{vzukovivc2019xy,
  title={XY model with antinematic interaction},
  author={{\v{Z}}ukovi{\v{c}}, Milan},
  journal={Physical Review E},
  volume={99},
  number={6},
  pages={062112},
  year={2019},
  publisher={APS}
}

@article{huhtinen2026stability2,
  author =        {Huhtinen, Kukka-Emilia},
  journal =       {arXiv preprint arXiv:2603.09954},
  title =         {Stability of flat-band Bose-Einstein condensation
                   from the geometry of compact localized states},
  year =          {2026},
}

@article{maimaiti2017compact,
	author = {Maimaiti, Wulayimu and Andreanov, Alexei and Park, Hee Chul and Gendelman, Oleg and Flach, Sergej},
	doi = {10.1103/PhysRevB.95.115135},
	journal = {Phys. Rev. B},
	month = {Mar},
	number = {11},
	pages = {115135},
	publisher = {American Physical Society},
	title = {Compact localized states and flat-band generators in one dimension},
	url = {https://link.aps.org/doi/10.1103/PhysRevB.95.115135},
	volume = {95},
	year = {2017}}

@article{maimaiti2021flatband,
 author = {Maimaiti, Wulayimu and Andreanov, Alexei and Flach, Sergej},
 title = {Flat-band generator in two dimensions},
 journal = {Phys. Rev. B},
 volume = {103},
 issue = {16},
 pages = {165116},
 numpages = {15},
 year = {2021},
 month = apr,
 publisher = {American Physical Society},
 doi = {10.1103/PhysRevB.103.165116},
 url = {https://link.aps.org/doi/10.1103/PhysRevB.103.165116},
}

@article{graf2021designing,
  title = {Designing flat-band tight-binding models with tunable multifold band touching points},
  author = {Graf, Ansgar and Pi\'echon, Fr\'ed\'eric},
  journal = {Phys. Rev. B},
  volume = {104},
  issue = {19},
  pages = {195128},
  numpages = {21},
  year = {2021},
  month = {Nov},
  publisher = {American Physical Society},
  doi = {10.1103/PhysRevB.104.195128},
  url = {https://link.aps.org/doi/10.1103/PhysRevB.104.195128}
}

@article{hwang2021general,
  title = {General construction of flat bands with and without band crossings based on wave function singularity},
  author = {Hwang, Yoonseok and Rhim, Jun-Won and Yang, Bohm-Jung},
  journal = {Phys. Rev. B},
  volume = {104},
  issue = {8},
  pages = {085144},
  numpages = {16},
  year = {2021},
  month = {Aug},
  publisher = {American Physical Society},
  doi = {10.1103/PhysRevB.104.085144},
  url = {https://link.aps.org/doi/10.1103/PhysRevB.104.085144}
}

\onecolumngrid
\section*{End Matter}
\twocolumngrid

\setcounter{equation}{0}
\renewcommand{\theequation}{E\arabic{equation}}

\setcounter{figure}{0}
\renewcommand{\thefigure}{E\arabic{figure}}

\sect{BdG spectra} 
Invoking Eq.~\eqref{eq:GS_fluctuation_spectrum} to the equation of motion \(i\partial_t p_l(t) = \partial \mc L_{\rm eff}/\partial p_l^*\),
the full BdG eigenvalue equation for the ABF lattice reads
\begin{align}
    \notag \lambda \chi_l &= \big[(\alpha_0-\mu)\chi_l+\alpha_-\chi_{l-1}+\alpha_+\chi_{l+1}\big] \\
    \notag &\qquad+\big(\beta_{l-}\Pi_{l-1}+\beta_{l0}\Pi_l+\beta_{l+}\Pi_{l+1}\big), \\
    \notag \lambda \Pi_l &= -\big[(\alpha_0-\mu)\Pi_l+\alpha_+\Pi_{l+1}+\alpha_-\Pi_{l-1}\big] \\
    \label{eq:homogeneous_bdg_particle_hole_real}
    &\qquad-\big(\beta_{l-}^*\chi_{l-1}+\beta_{l0}^*\chi_l+\beta_{l+}^*\chi_{l+1}\big),
\end{align}
where
\begin{align}
    \notag \alpha_0 &= g\big(2c^4|a_l|^2 + 2s^4|a_{l-1}|^2 + 2s^2c^2\big(|b_l|^2 + |b_{l-1}|^2\big)\big),\\
    \notag \alpha_{+} &= -2gs^2c^2\big(|a_l|^2 - |b_l|^2\big), \\
    \notag \alpha_{-} &= -2gs^2c^2\big(|a_{l-1}|^2 - |b_{l-1}|^2\big),  \\
    \notag \beta_{l0} &= g\big(c^4a_l^2 + s^4a_{l-1}^2 + s^2c^2\big(b_l^2 + b_{l-1}^2\big)\big), \\
    \notag \beta_{l+} &= -gs^2c^2\big(a_l^2 - b_l^2\big),\\
    \label{eeq:bdg_ev}
    \beta_{l-} &= -gs^2c^2\big(a_{l-1}^2 - b_{l-1}^2\big).
\end{align}
Here, \(a_l\) and \(b_l\) denote the ground state amplitudes, with \(f_l=0\) from     Eq.~\eqref{eq:anbn_in_terms_of_pnfn}.
Note that the entire eigenvalue equation Eq.~\eqref{eq:homogeneous_bdg_particle_hole_real}
 scales linearly with \(gn\) which is expected.

This eigenvalue equation has effective disorder from the nematic phase choices for \(\theta > \pi/8\).

\begin{figure}[t]
    \centering
    \includegraphics[trim={0.08cm 0.2cm 0.03cm 0.12cm}, clip,width=\linewidth]{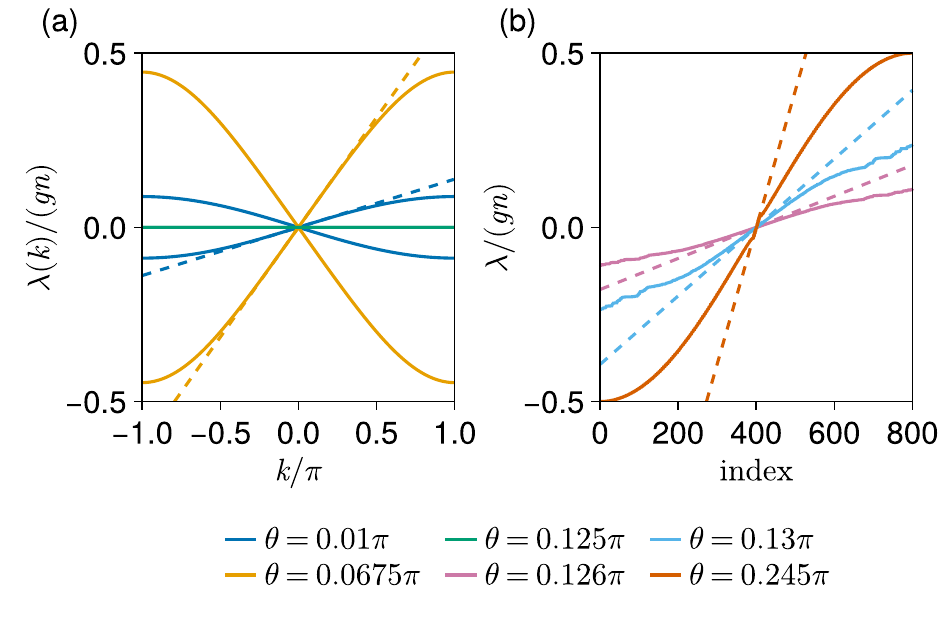}
    \caption{
        Full BdG spectra of ABF GP. (a) \(\lambda(k)\) vs. \(k\) for \(0 < \theta < \pi/8\) and (b) \(\lambda\) for \(\pi/8 < \theta < \pi/4\) sorted in increasing order. 
        The dashed lines are analytical computations of speed of sound \(c_s\).
    }
    \label{fig:abf_bdg}
\end{figure}

In Fig.~\ref{fig:abf_bdg}, panels (a) and (b) show the BdG spectra for \(0 \le \theta \le \pi/8\) and \(\pi/8 \le \theta < \pi/4\), respectively, excluding \(\theta=\pi/4\), computed numerically using exact diagonalization from representative ground states obtained above.
The solid lines show the numerical BdG spectra, while the dashed lines indicate the analytically predicted low-energy dispersion with slope \(c_s\). 
For \(\theta > \pi/8\), the BdG spectrum no longer respects translational symmetry, since it is defined on a symmetry-broken background with random phases \(\Delta \phi_l\). 
Nevertheless, this effective disorder is negligible in the low-energy sector, as manifested by the configuration-independent stiffness and sound velocity.
Assigning a momentum spacing \(\Delta k = 2\pi/L\) to the mode index, we find that the low-energy dispersion is in excellent agreement with the sound velocity predicted by the hydrodynamic theory in the Main text [Eqs.~\eqref{eq:stiffness}; see also
See SM~\cite{supp_local}].
As explained in the Main Text, at \(\theta=\pi/8\), the entire BdG spectrum vanishes completely.

\sect{Ground state selection via order-by-disorder}

\begin{figure}[tb]
    \centering
    \includegraphics[trim={1cm 3cm 0.2cm 0.8cm}, clip, width=0.98\linewidth]{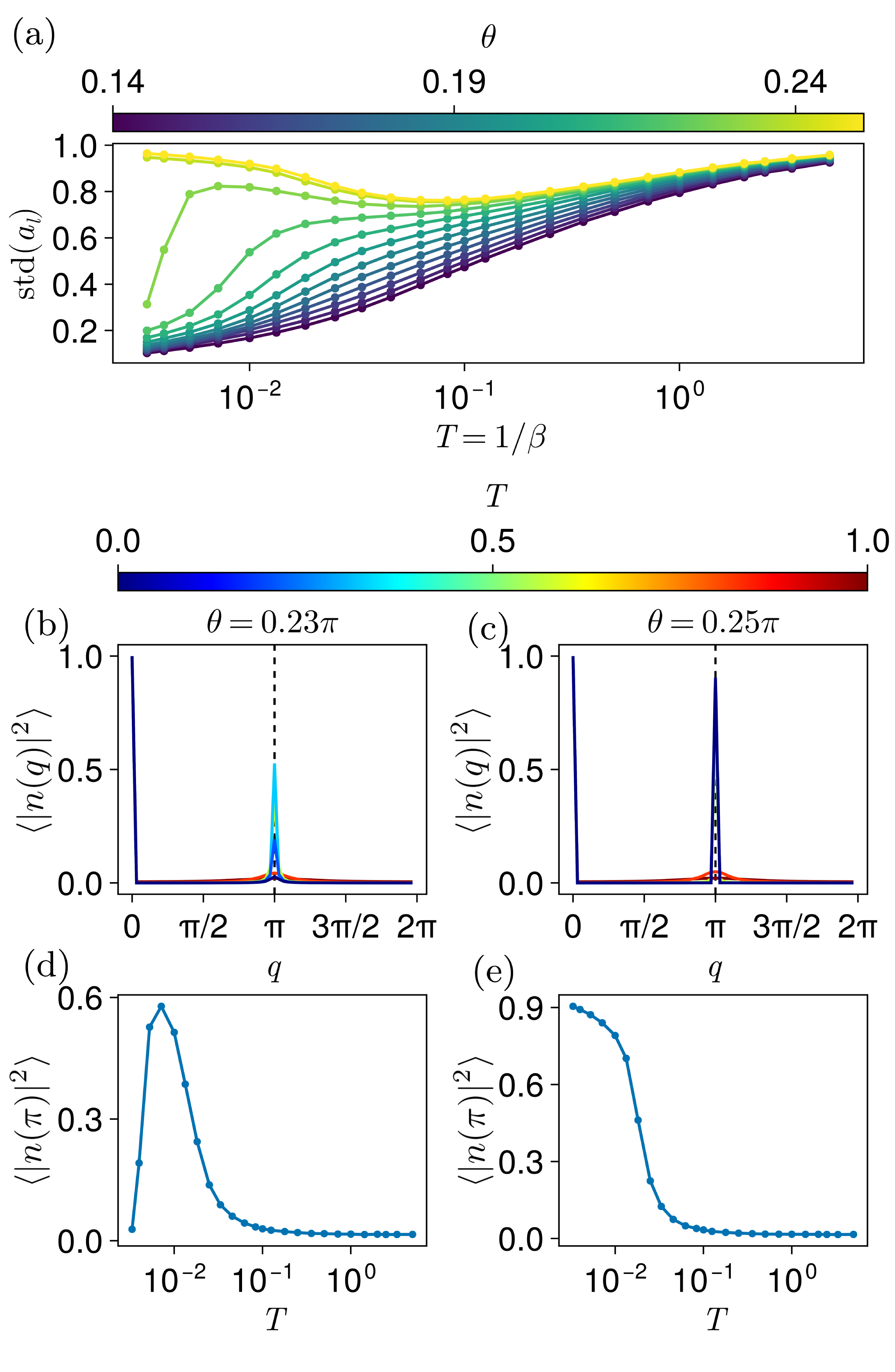}
    \caption{(a) Ensemble-averaged spatial density fluctuation, \(\mathrm{std}(a_l)\), versus temperature.
    (b)-(c) Density structure factor, \(\langle |n(q)|^2 \rangle\), for \(\theta/\pi = 0.23\) and \(0.25\) at different temperatures (colors).
    (d)-(e) Temperature dependence of the peak value at \(q = \pi\).}
    \label{sfig:density_modulation}
\end{figure}

\begin{figure}[ht!]
    \centering
    \includegraphics[trim={0 0.4cm 0 0}, clip, width=0.98\linewidth]{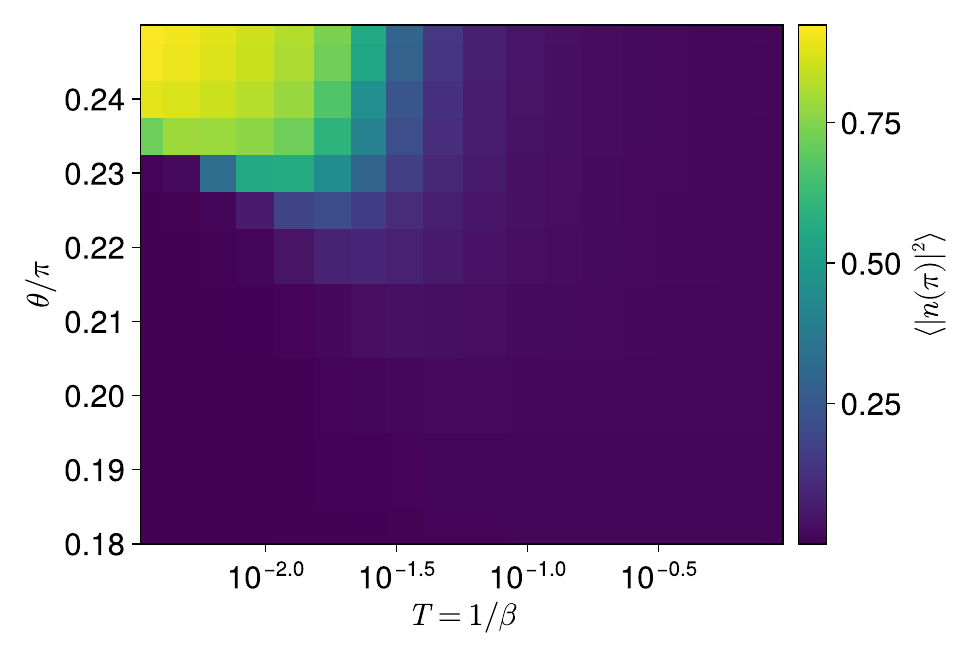}
    \caption{Density-modulation crossover diagram obtained from the density structure factor at \(q=\pi\), \(\langle |n(\pi)|^2\rangle\), as a function of temperature \(T=1/\beta\) and \(\theta/\pi\).}
    \label{sfig:density_crossover_diagram}
\end{figure}

Let us remind that the tiling ground states at \(\theta = \pi/4\) are homogeneous density states in the original lattice, but modulated density states in the projected FB basis.
At variance, the nematic ground states are homogeneous density states in both the original lattice and the projected FB basis, for the entire interval \( \pi/8 \leq \theta \leq \pi/4\). 
In the following we work in the projected FB basis in order to distinguish the tiling states from the nematic ones.

We have obtained the finite temperature samples down to \(T = 0.001\), using the same algorithms for obtaining the GS, but without gradient descent fine optimization this time.
In Fig.~\ref{sfig:density_modulation}(b,c), we show the ensemble-averaged Fourier transform of the norm densities \(n_l = |p_l|^2\), denoted \(\langle |n(q)|^2\rangle\) (structure factor of the density), and the corresponding peak value at \(q=\pi\) for \(\theta/\pi = 0.23\) and \(0.25\) at finite temperatures. 
The peak at \(q=\pi\), whose magnitude becomes comparable to that at \(q=0\), indicates a strong tendency toward density modulation, consistent with the order-by-disorder mechanism discussed above. 

Interestingly, even when the tiling states are no longer exact ground states away from \(\theta=\pi/4\), a remnant of this mechanism persists. 
We find that there exists a finite-temperature window in which the tiling states are favored, before the system eventually crosses over to the nematic ground state at lower temperature as shown in Fig.~\ref{sfig:density_modulation}(d,e).
Thus, near but not exactly at \(\theta = \pi/4\), the crossover towards the tiling states is observed at finite temperatures through the CLS density order parameter \(\langle | n(\pi) |^2\rangle \).
Figure~\ref{sfig:density_crossover_diagram} summarizes this crossover through the density structure factor at \(q=\pi\), showing that the density-modulated response extends over a finite region near \(\theta=\pi/4\).

\newpage

\foreach \x in {1,...,17}
{%
\clearpage
\includepdf[pages={\x}]{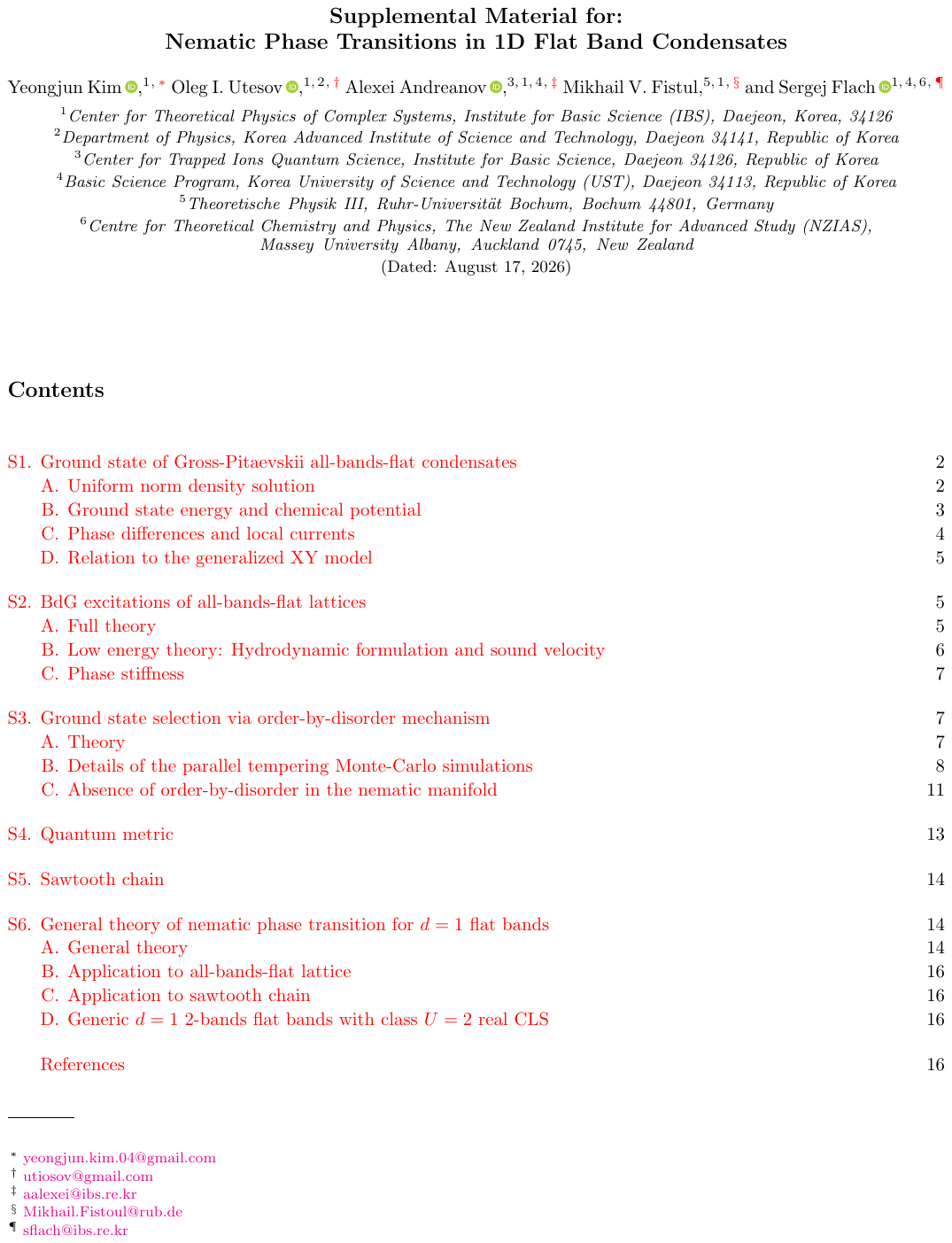}
}

\clearpage
\pagestyle{empty}
\end{document}